\documentclass[12pt,a4paper]{article}
\usepackage[utf8]{inputenc}
\usepackage{amsmath}
\usepackage{amsfonts}
\usepackage{amssymb}
\usepackage{mathtools}
\usepackage{graphicx}
\usepackage{cite}
\graphicspath{ {./figures/} }
\usepackage[left=2cm,right=2cm,top=2cm,bottom=2cm]{geometry}

\newcommand{\Ykw}{\bar{Y}}

\author{Tetiana Obikhod, Ievgenii Petrenko}
\title{Searches for heavy Higgs bosons in the framework of 2HDM model}
\date{%
    Institute for Nuclear Research NAS of Ukraine, Kyiv 03028, Ukraine\\%
    \today
}
\begin{document}

\maketitle

\section{Abstract}

	The searches for heavy neutral and charged Higgs bosons are performed through the calculations of production cross sections using MadGraph5aMC@NLO program with ansatz of Yukawa coupling and the restricted parameter space connected with LHC Run 2 data. The searches for heavy resonances are performed in the framework of 2HDM model over the mass range 0.1–1 TeV for the  $p p \to A t\overline{b}$, $p p \to H^+b\overline{t}$ , $p p\to H^+t\overline{t}$ , $p p\to H H Z$ decay modes. The presented data demonstrate the jump in the production cross section of $H^+ b \overline{t}$ and $H H Z$ production processes in the mass range of 100-200 GeV and  100-300 GeV accordingly at energy of 14 TeV.

\section{Introduction}

The experimental largest and most powerful collider – LHC \cite{Lincoln:2009zz} was built generally for searches of new physics beyond Standard Model (SM). 10 years of LHC physics are associated with 280 petabytes of data, obtaining of about 8 millions of Higgs bosons and with more than 50 computing collaborations. And in spite of these facts "LHC has gathered just 1 percent of the total amount of data it aims to collect" – said theorist Nima Arkani-Hamed in 2016 year. With that data, scientists can indeed rule out the most vanilla form of supersymmetry. Difficulties associated with the multiplicity of particle production in proton collisions and with theoretical description of hadronization slow down the processing of data, but lead to assumptions about a new physics in studying the angles of escape and energy of jets. Such physics is described by different models, which could explain difficulties of SM, one of the most important of which is connected with vacuum properties – the hierarchy problem. Supersymmetry (SUSY) – is the theory, which solves the problem.
\par
One of the most interesting SUSY models is 2-Higgs-Doublet Model (2HDM) \cite{BRANCO20121}, which uses suppressed Yukawa couplings for the explanation of the following experimental data:
\begin{itemize}
	\item neutrino oscillations;
	\item dark matter candidate;
	\item CP-violation;
	\item mass of Higgs boson.
\end{itemize}
\par
	In the late 2-3 years there are appeared experimental papers connected with searches for Higgs bosons in the mass range (70-110) GeV \cite{2019320}, (200-250) GeV \cite{Aad:2020zxo}, (400 – 700) GeV 
	\cite{CMS} with significance from 1.9 to 4 standard deviations. The accumulation and analysis of experimental data at higher energies should shed light on the existence of the extended Higgs boson sector to test consistence with 2HDM interpretation. 
\par
The purpose of our paper is to study the extended Higgs boson sector using computer modeling and experimentally agreed scenarios of 2HDM parameter space.

\section{The review of Two-Higgs-Doublet Model}

The 2HDM is presented by the Higgs potential and the Yukawa couplings of the two scalar-doublets to the three generations of quarks and leptons. The gauge-invariant renormalizable Higgs scalar potential is given by

\begin{multline*}
V  = m^2_{11} \Phi_1^{\dag}\Phi_1 + m^2_{22}\Phi_2^{\dag}\Phi_2 - \left(m^2_{12} \Phi^{\dag}_1\Phi_2 + h.c. \right) \\
 + \frac{1}{2}\lambda_1 \left( \Phi_1^\dag\Phi_1 \right)^2
 + \frac{1}{2}\lambda_2 \left( \Phi_2^\dag\Phi_2 \right)^2 
 + \lambda_3 \left( \Phi_1^\dag\Phi_1 \right) \left( \Phi_2^\dag\Phi_2 \right)
 + \lambda_4 \left( \Phi_1^{\dag}\Phi_2 \right)\left( \Phi_2^{\dag}\Phi_1 \right) \\
 + \left[ \frac{1}{2}\lambda_5 (\Phi_1^{\dag}\Phi_2)^2  + \left[ \lambda_6 (\Phi_1^\dag\Phi_1) + \lambda_7 (\Phi_2^\dag\Phi_2) \right]\Phi^{\dag}\Phi_2 + h.c. \right]
\end{multline*}
where $m^2_{11}$, $m^2_{22}$ and $\lambda_1$, ..., $\lambda_4$ are real parameters. In general $m^2_{12}$, $\lambda_5$,  $\lambda_6$ and  $\lambda_7$ are complex.
$\lambda_i$, for $i$ = 1 to 7, are all the Higgs quartic couplings. After the spontaneous breaking of the EW symmetry, five physical Higgs particles are left in the spectrum: one charged Higgs pair, $H^{\pm}$, one CP-odd neutral scalar, $A$ and two $CP$-even neutral states, $H$ and $h$.
The components $\Phi_i(i=1,2)$ denote two complex SU(2) doublet scalar fields:

\begin{equation*}
\Phi_i = \left( \stackrel{\phi_i^+}{\frac{\upsilon_i + \phi_i + i \chi_i}{\sqrt{2}}} \right) 
\end{equation*}

The physical Higgs bosons are received from two Higgs doublets $H_1$ and $H_2$ in the basis $\Phi_1$ and $\Phi_2$
\begin{equation*}
\begin{split}
H_1 = \Phi_1 \cos \beta + e^{-i\xi}\Phi_2 \sin\beta\\
H_1 = -e^{-i\xi} \Phi_1 \sin \beta + \Phi_2 \cos\beta.
\end{split}
\end{equation*}
\par
The 2HDM parameter space is described by six free parameters: the physical Higgs masses ($m_h$, $m_H$, $m_A$ and $m_{H^\pm}$), the mixing angle between the two CP-even Higgses ($\alpha$), and the ratio of the two vacuum expectation values ($\tan\beta = \upsilon_2/\upsilon_1$)
\par
For simplicity, we have set the phase $\xi$ to be zero.
So, after SUSY breaking we have five Higgs bosons expressed by the following formulas:
\begin{equation*}
\begin{split}
H^\pm = -\sin\beta\phi^\pm_1 + \cos\beta\phi^\pm_2\\
A = -\sin\beta\chi_1 + \cos\beta\chi_2\\
H = \cos\alpha\phi_1 + \sin\alpha\phi_2\\
h = -\sin\alpha\phi_1 + \cos\alpha\phi_2
\end{split}
\end{equation*}
The ratio between the angles $\alpha$, $\beta$ leads to different SM-like Higgs bosons. For $\beta - \alpha \sim 90^\circ$, the lightest CP even Higgs boson has couplings like SM-Higgs, h; for $\beta - \alpha \sim 0^\circ$, the lightest SM-like Higgs boson is H. 

\begin{equation*}
H^{SM} = h\sin (\alpha - \beta) - H \cos (\alpha - \beta).
\end{equation*}

The possibility of tree level flavour-changing neutral currents (FCNC), leads to interaction of Higgs doublets with quarks (f) and leptons (l) with not flavour diagonal Yukawa couplings. The Yukawa Lagrangian with parameters  is the following

\begin{multline*}
\mathcal{L}^{\text{2HDM}}_{\text{Yukawa}} = - \sum_{f = u,d,l} \frac{m_f}{\upsilon}\left( \xi^f_h\overline{f}fh + \xi^f_H\overline{f}fH
 - i\xi^f_A\overline{f}\gamma_5 fA \right) \\
	- \left\lbrace \frac{\sqrt{2}V_{ud}}{\upsilon}\overline{u} \left( m_u\xi^u_AP_L  
	+ m_d\xi^d_AP_R \right)d\text{H}^+
	 + \frac{\sqrt{2}m_1\xi^l_A}{\upsilon}\overline{\nu}_L l_R H^+ + h.c.  \right\rbrace,
\end{multline*}
where $P_{L/R}$ are the projection operators for left-/right-handed fermions, and $V_{ud}$ denotes the appropriate element of the CKM matrix. Factors $\xi$ are presented in table \ref{table:xi}.
\begin{table}[t]
\centering
\begin{tabular}{|c|c|c|}
\hline 
 & Type I & Type II \\ 
\hline 
$\xi^u_h$ & $\cos\alpha/\sin\beta$ & $\cos\alpha/\sin\beta$ \\ 
\hline 
$\xi^d_h$ & $\cos\alpha/\sin\beta$ & $-\sin\alpha/\cos\beta$ \\ 
\hline 
$\xi^l_h$ & $\cos\alpha/\sin\beta$ & $-\sin\alpha/\cos\beta$ \\ 
\hline 
$\xi^u_H$ & $\sin\alpha/\sin\beta$ & $\sin\alpha/\sin\beta$ \\ 
\hline 
$\xi^d_H$ & $\sin\alpha/\sin\beta$ & $\cos\alpha/\cos\beta$ \\ 
\hline 
$\xi^l_H$ & $\sin\alpha/\sin\beta$ & $\cos\alpha/\cos\beta$ \\ 
\hline 
$\xi^u_A$ & $\cot\beta$ & $\cot\beta$ \\ 
\hline 
$\xi^d_A$ & $-\cot\beta$ & $\tan\beta$ \\ 
\hline 
$\xi^l_A$ & $-\cot\beta$ & $\tan\beta$ \\ 
\hline 
\end{tabular} 
\caption{Yukawa couplings $u$, $d$, $l$ to the neutral Higgs bosons, $h$, $H$, $A$}\label{table:xi}
\end{table}
\par
As we will do the calculations for two models of 2HDM we presented Yukawa couplings only for Type I and Type II models.

\section{Benchmark scenarios within the 2HDM model}

There have been many theoretical and experimental studies searching for additional scalar particles connected with significant constraints on the parameter space in the framework of the 2HDM \cite{Sanyal:2019xcp, Babu:2018uik}. Different decay modes and coupling measurements provide additional search channels in the parameter regions and exclusion regions obtained from the 13 TeV by CMS and ATLAS collaborations \cite{Kling:2020hmi}. 
\par
The new parameter space of the model is spanned by the three new Yukawa couplings, by the mass of the heavy neutral Higgs boson $H$ and by the mixing angle $\alpha - \beta$

\begin{equation*}
\lbrace  \Ykw_t, \Ykw_b, \Ykw_\tau, M_H, \sin(\alpha-\beta) \rbrace
\end{equation*}

\par
Using simple ansatz for the Yukawa couplings \cite{Babu:2018uik} with large deviation from SM we considered three benchmark scenarios (BP1, BP2, BP3) within the 2HDM model presented in table \ref{table:BP}.

\begin{table}
\centering
\begin{tabular}{|c|c|c|c|c|c|}
\hline 
Benchmark points & $\Ykw_t$ & $\Ykw_b$ & $\Ykw_\tau$ & $\sin(\alpha-\beta)$ & $M_H$(GeV) \\ 
\hline 
BP1 & +1.01 & -0.10 & $10^{-3}$ & +0.50 & 500 \\ 
\hline 
BP2 & -1.0 & +0.01 & $10^{-3}$ & -0.1 & 600 \\ 
\hline 
BP3 & 1.25 & +0.05 & $10^{-3}$ & -0.2 & 680 \\ 
\hline 
\end{tabular} 
\caption{Sample points on parameter space}\label{table:BP}
\end{table}

\section{Calculations of Higgs boson production cross sections }

To probe the extended Higgs sector we used the known methods connected with the modifications of the SM-like Higgs couplings \cite{Heinemeyer:2013tqa}, and direct searches \cite{Heinemeyer:1559921}. In this part we presented the production cross section calculations within 2HDM model using the additional couplings of the top and bottom quarks, which lead to distinct signatures in the hh and qq~h production rates. Our purpose is to examine some deviations in the properties of SM predictions within experimental limits. For this purpose we used MadGraph5aMC@NLO program \cite{Alwall:2014hca} with NLO calculations of production cross sections presented below. It gives us the possibility to consider 2HDMtII\_NLO model available for 2HDM – I or II models with corresponding $\alpha$, $\beta$ angles. 

\subsection{The modifications of the SM-like Higgs couplings}
We considered processes for $A$ and $H^+$ boson production because of the presence of top and bottom Yukawa couplings assuming SM-like branching fractions of the Higgs bosons. Di-Higgs production $pp \to H H Z$ is correlated with the resonance mass and can be much larger than the SM value. 
\par
\begin{figure}[htbp]
\minipage{0.48\textwidth}
   \includegraphics[width=\linewidth]{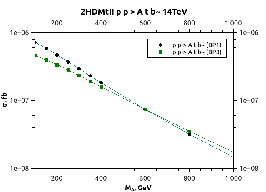}
 \caption{\label{fig:Atb} Production cross sections of $A$ boson as a function of its mass $M_A$}
\endminipage\hfill
\minipage{0.48\textwidth}
 \includegraphics[width=\linewidth]{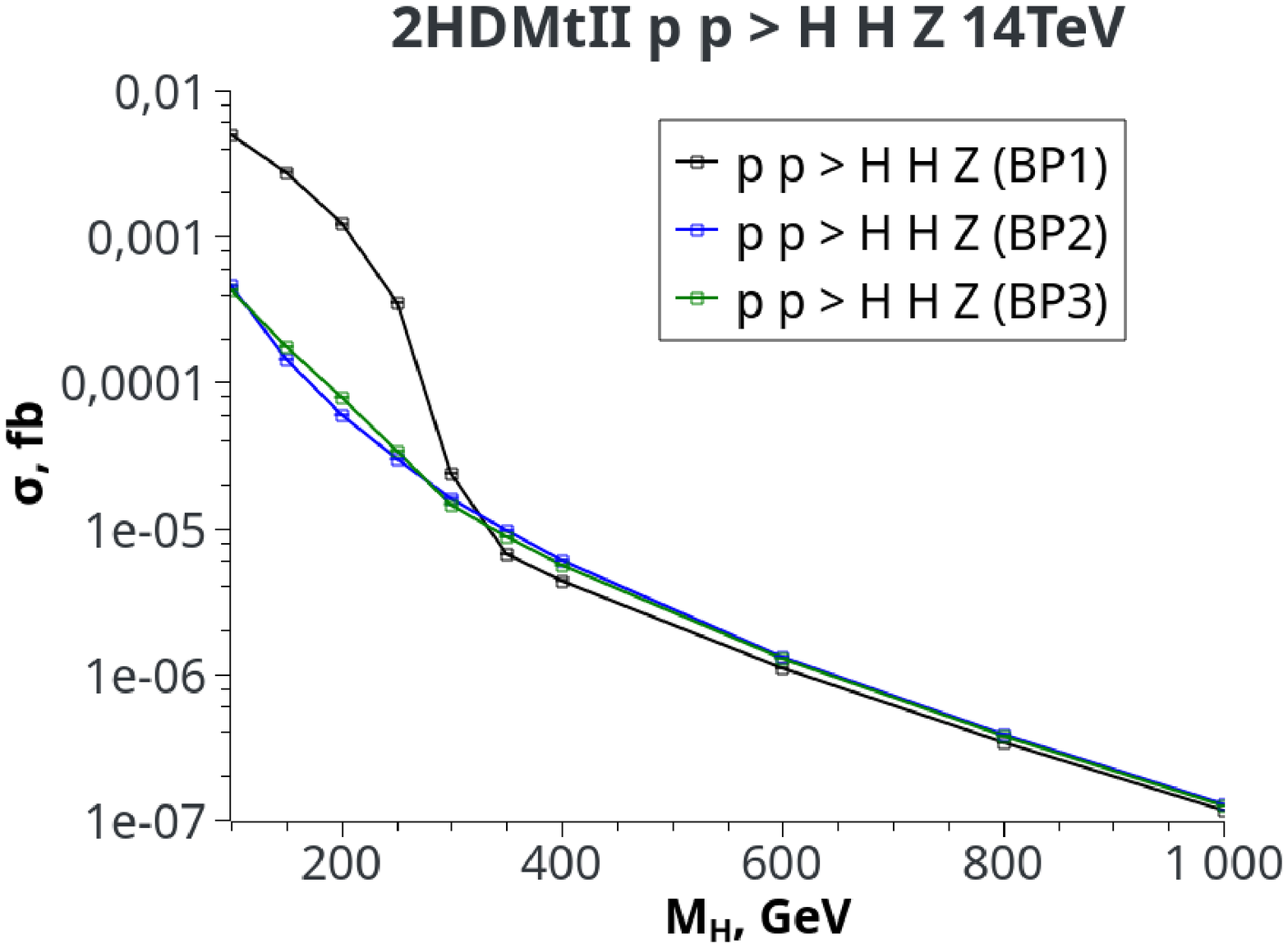}
 \caption{\label{fig:HHZ} Production cross section of H boson as a function of mass $M_H$}
\endminipage\hfill
\end{figure}
\begin{figure}[htbp]
 \includegraphics[width=0.45\textwidth]{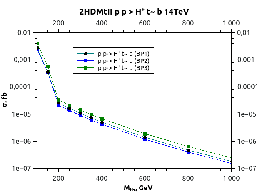}
 \includegraphics[width=0.45\textwidth]{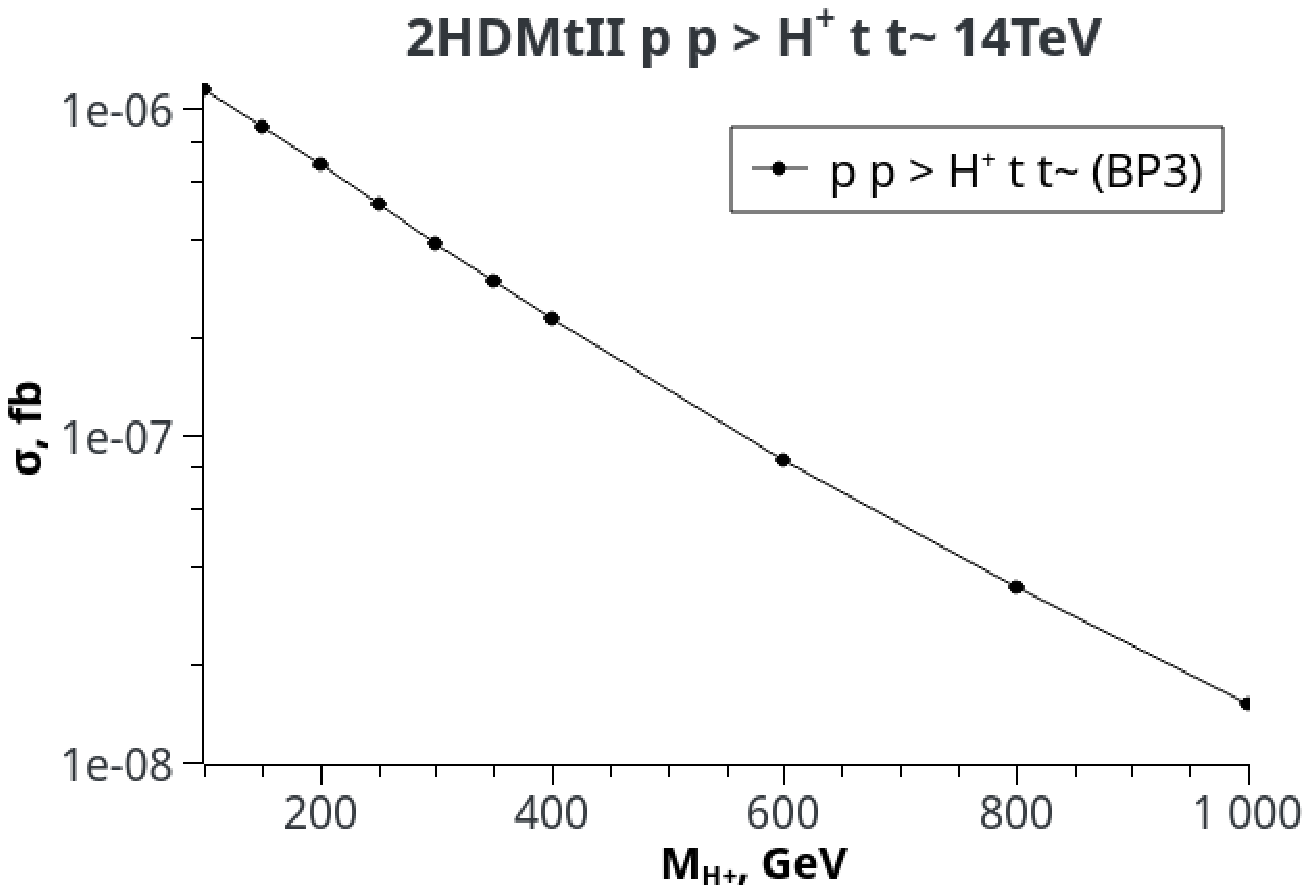}
 \caption{\label{fig:Htb} Production cross sections of $H^+$ boson as a function of its mass, $M_{H^+}$ for left: $pp \to H^+b\overline{t}$ process;  right: $pp \to H^+t\overline{t}$ process  }
\end{figure}

We did the calculations using new ansatz for the Yukawa couplings of the Higgs doublets proposed in \cite{Babu:2018uik} and presented in table \ref{table:xi}. For this purpose we considered the following processes at energy of 14 TeV:
\begin{itemize}
	\item $pp \to At\overline{b}$, Fig.\ref{fig:Atb};
	\item $pp \to H H Z$, Fig.\ref{fig:HHZ};
	\item $pp \to H^+b\overline{t}$, $pp\to H^+t\overline{t}$, Fig.\ref{fig:Htb};
\end{itemize}

\par
The considered processes shows sharp increase in the production cross section of H+bt and HHZ production processes in the mass range of 100-200 GeV and  100-300 GeV accordingly. We also considered $H^+t\overline{t}$ production process, as its rate is proportional to the square of the top Yukawa coupling with the Higgs boson, which is disrupted in the 2HDM model. From Fig.\ref{fig:Htb} (right part) we see the significant suppression of this process. As for the A boson we didn’t see any sharp jumps in cross section behavior in the considered mass range 100 GeV – 1000 GeV at energy of 14 TeV.

\subsection{Direct searches for BSM Higgses}
The window for searches of heavy BSM Higgses is connected with decay modes for comprehensive examination of the current constraints on the 2HDM parameter space. We summarized the latest LHC searches, \cite{Kling:2020hmi} together with the interpretation of these results for the Type-I and Type-II 2HDM model. Due to the very stringent restrictions on the parameter space obtained from the experimental data 
\cite{Chen:2018shg,Gu:2017ckc,Chen:2019pkq,Chen:2019rdk,Chen:2019bay,Su:2019ibd,Kling:2018xud} and inclusion at the LHC searches of both search channels related to the coupling measurements and the decay channels of the Higgs boson, we selected only three scenarios with points that have not yet been covered by the experimental exceptions for 2HDM (I or II). At the same time, we took one scenario that is closest to the SM one, ($\sin(\beta-\alpha)=0.99$, $\tan\beta=2$), for checking deviations from the SM of other calculations according to the restrictions of the experimental parameter space.
\par
\begin{figure}[htbp]
\minipage{0.48\textwidth}
   \includegraphics[width=\linewidth]{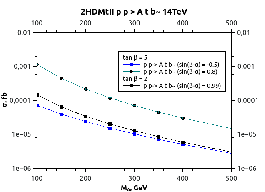}
 \caption{\label{fig:Atb-sin} Production cross sections of $A$ boson as a function of its mass, $M_A$}
\endminipage\hfill
\minipage{0.48\textwidth}
 \includegraphics[width=\linewidth]{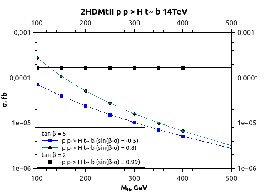}
 \caption{\label{fig:Htb-sin} Production cross section of H boson as a function of mass, $M_H$}
\endminipage\hfill
\end{figure}

Using MadGraph5aMC@NLO program we calculated production cross sections of the processes at 14 TeV:
\begin{itemize}
	\item $pp \to A t\overline{b}$, Fig.\ref{fig:Atb-sin};
	\item $pp \to H b\overline{t}$, Fig.\ref{fig:Htb-sin};
\end{itemize}

The parameter spaces are presented at the top of the corresponding figures. For comparison with SM model we presented the calculations at alignment limit, $\sin\left(\beta - \alpha \right) \sim 1$ with consistent couplings of Higgs to bosons and fermions predicted by SM. 
\par
In Fig.\ref{fig:Atb-sin} is clearly presented the importance of the $\tan\beta $ value, because of the correlation between $\sin\left(\beta - \alpha \right)$  and $\tan\beta$. From the comparison of the production cross sections we see the larger value for $\sin\left(\beta - \alpha \right)=0.8$ ($\tan\beta =5$) compared to $\sin\left(\beta - \alpha \right)=0.99$ ($\tan\beta =2$) in spite of the nearest to alignment limit values ($\sin\left(\beta - \alpha \right) \sim 1$) of two scenarios. The value $\sin\left(\beta - \alpha \right)=-0.5$ was taken from the parameter space not measured before.
\par
In Fig.\ref{fig:Htb-sin} we also didn't see any sharp burst of peak in production cross section of Higgs boson in 2HDM model. For the investigation of the deviation from SM we presented calculations at about $\sin(\beta-\alpha)\sim 1$ , which compatible with SM predictions. From the comparison of three scenarios we see that only the value of $\sin(\beta-\alpha)=0.8$ ($\tan\beta=5$) present some excess in the production cross section for the mass range of 100-120 GeV.

\section{Conclusions}

The searches for BSM physics at the LHC are associated generally with SUSY searches. As superpartners of SM Higgs boson presented in 2HDM model are the lightest supersymmetric particles, their searches are the most optimal one due to the expected mass of the extended Higgs boson sector in the mass region up to 1 TeV. Latest experimental data are related to the searches for additional Higgs bosons in the mass range of about 95-100 GeV, 350-400 GeV and 600-650 GeV. These searches are accompanied by severe restrictions on the parameter space. So our purpose was to take the most comprehensive parameter restrictions connected with ansatz for the Yukawa couplings and with the latest experimental parameter space from decay channels and Higgs coupling measurements. 
\par
Our calculations were divided into two parts. The calculations according to the first part connected with modification of Yukawa couplings demonstrated the clear and bright jump in production cross section of $H^+b\overline{t}$ and $HHZ$ production processes in the mass range of 100-200 GeV and  100-300 GeV accordingly at energy of 14 TeV. So, we can say about the possibility to find charged $H^+$ and CP-even $H$ bosons in the corresponding mass ranges. As for the second part, we didn’t see any essential deviations from the SM. 
\par
The consequence of our calculations is the conclusion about the predominance of Yukawa coupling modifications over the modifications of the angles $\alpha$, $\beta$ in the searches for SUSY signal.

\bibliographystyle{ieeetr}

\end{document}